\documentclass[aip, numerical,reprint,floatfix]{revtex4-1}
\usepackage[dvips]{graphicx}
\usepackage{color}

\begin{document}


\title{
Innovative Demodulation Scheme for Coherent Detectors in CMB Experiments} 



\author{K. Ishidoshiro}
\altaffiliation[Present address: ]
{Research Center for Neutrino Science, Tohoku University, Sendai 980-8578, Japan}
\email{koji@awa.tohoku.ac.jp}
\author{Y. Chinone}
\author{M. Hasegawa}
\author{M. Hazumi}
\author{M. Nagai}
\author{O. Tajima}
\affiliation{Institute of Particle and Nuclear Studies, High Energy Accelerator Research Organization~(KEK), 
Oho, Tsukuba, Ibaraki, 305-0801, Japan}


\date{\today}

\begin{abstract}
We propose an innovative demodulation scheme for coherent detectors used 
in cosmic microwave background polarization experiments. 
Removal of non-white noise, e.g., narrow-band noise, in detectors 
is one of the key requirements for the experiments. 
A combination of modulation and demodulation is used to extract polarization signals as well as to suppress such noise. 
Traditional demodulation, which is based on the two-point numerical differentiation, 
works as a first-order high pass filter for the noise.
The proposed demodulation is based on the three-point numerical differentiation.
It works as a second-order high pass filter.
By using a real detector, we confirmed significant improvements of suppression power for the narrow-band noise.
We also found improvement of the noise floor.
\end{abstract}

\maketitle 


Detection of primordial gravitational waves could provide 
a new and unique window on the very early universe. 
Degree-scale odd-parity patterns in cosmic microwave background (CMB) polarization, 
$B$-modes, are a smoking-gun signature of the primordial gravitational waves~\cite{intro1}. 
Since $B$-modes are very faint ($\lesssim 100$~nK), 
it is essential to accumulate observation data for long time, e.g., a few years.
We evaluate the $B$-modes by using the CMB power spectrum~\cite{Durrer:2008eb}.
Suppose polarization sensitive detectors have white Gaussian noise, 
the precision of the measured spectrum is inversely proportional to the data integration time.
However, for real experiments,
the data contain non-white components in analysis band; line noise (narrow band noise). 
Potential sources of such noise are AC line noise, mechanical resonances of the telescope and so on. 
The line noise degrades the sensitivity improvement with respect to data integration time. 
In the case of the experiments which use coherent detectors~\cite{QUIET,PLANK}, 
the CMB polarization signal is extracted by a combination of modulation and demodulation.
This technique can also suppress the non-white noise. 

In this paper, we suggest improved demodulation scheme based on the idea extension to three-point differentiation from the two-point one. 
It drastically improves the suppression of the line noise in the analysis band. 
Therefore, the residual line noise in the two-point demodulation can be suppressed further, 
even if we do not know the sources of the line noise. 

In the case of coherent detectors, 
the phase of the input CMB polarization signal is shifted  
by changing two microwave paths which have different path lengths. 
The difference is designed to correspond to a half wave length of the input CMB signals. 
Therefore, periodic path switching works as sign modulation at a given frequency. 
Two-point differentiation between two modulation states, namely two-point demodulation, 
extracts the polarization signals and filters noise components at frequencies lower
than the modulation frequency. 
As a result, the demodulated data have a white spectral density 
whose floor is determined by the input noise at around the modulation frequency. 

The raw output at time $t_i$ of the detector can be described as follows:
\begin{equation}\label{eq1}
 S_i = a_i P_i + N_i,
\end{equation}
where $a_i$ is the phase state of the polarization signals,
$P_i$ and $N_i$ are the magnitude of the polarization signals and the noise respectively.
Here, $i$ is the normalized time index with respect to a half cycle of the modulation frequency.
We re-define $a_i = +1 (-1)$ in case the $i$ is odd (even) number;
\begin{eqnarray}
 S_{2n+1} &=& +P_{2n+1} + N_{2n+1},
	\\
 S_{2n~~~} &=& -P_{2n~~~} + N_{2n}, 
\end{eqnarray}
where $n$ is the integer number, i.e., $n= (0, 1, 2, ...)$.

\begin{figure}[htb]
\centering
 \includegraphics[width=6.5cm]{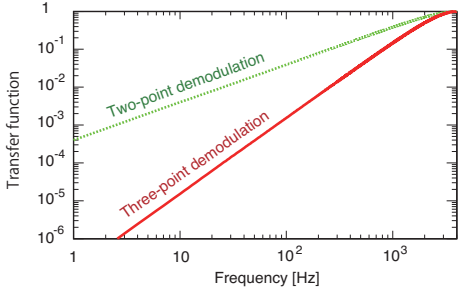}
 \caption{Transfer functions which indicate the power of noise filtering: two-point demodulation (dotted curves) 
and three-point demodulation (solid curves) as a function of frequency. 
The two-point demodulation works as a first-order high pass filter. 
The three-point demodulation works as a second-order high pass filter.
}\label{fig-tf}
\end{figure}

Traditional demodulation, namely two-point demodulation, can be defined as,
 \begin{eqnarray}
  D^{2p} &\equiv& \frac{ S_{2n+1} - S_{2n} }{2}, 
 	\\
 	&=& \frac{ P_{2n+1} + P_{2n} }{2} + \frac{N_{2n+1} - N_{2n}}{2}.
 \end{eqnarray}
Here $\{2(t_{2n+1} - t_{2n})\}^{-1}$ corresponds to the modulation frequency. 
In case of the QUIET experiment~\cite{QUIET}, 
the modulation frequency is $4$~kHz~\cite{footnote}  
to suppress the $1/f$ noise because its knee frequency is typically a few kHz.
Within such a short time interval, variation of observing direction, i.e., variation of the CMB signals, is negligible ($P_{2n} = P_{2n+1}$).
The two-point demodulation is re-written as follows,
\begin{eqnarray}
 D^{2p}_k &=& P_{k} + \Delta N_k^{2p},
 \end{eqnarray}
 where $k$ is the time index which synchronizes with the modulation, 
i.e., $t_k \equiv (t_{2n+1} + t_{2n})/2$ 
and $\Delta N_k^{2p} \equiv ( N_{2n+1} - N_{2n})/2$ is the two-point difference of the noise. 
The formula indicates that the two-point modulation works as a first-order high pass filter for the noise. 
The transfer function from $N_k$ to $\Delta N_k^{2p}$, i.e., noise filtering power, is shown as a function of frequency in Fig.~\ref{fig-tf}.

Three-point demodulation improves suppression power further by applying one more high pass filter. 
It is defined as follows,
\begin{eqnarray}
 D_l^{3p} &\equiv&  
 	\frac{1}{2} \Biggl[ \frac{ S_{2n+1} - S_{2n} }{2} - \frac{ S_{2n} - S_{2n-1} }{2} \Biggr]
 	\\
 	 &=& 
  	\frac{1}{2} \Biggl[ \frac{ P_{2n+1} + P_{2n} }{2} + \frac{ P_{2n} + P_{2n-1} }{2} \Biggr] \nonumber
 	\\
 	&&
 	+ 
  	\frac{1}{2} \Biggl[ \frac{ N_{2n+1} - N_{2n} }{2} - \frac{ N_{2n} - N_{2n-1} }{2} \Biggr]	
  	\\
 	&=& P_{l} + \Delta N_{l}^{3p},
 \end{eqnarray}
where $\Delta N_{l}^{3p} \equiv \left[ (N_{2n+1} - N_{2n})/2  - (N_{2n} - N_{2n-1})/2 \right]/2$ and $l \equiv 2n$.
We can also treat $P_l$ does not vary within the time interval between $2n-1$ and $2n+1$. 
The term  $\Delta N_{l}^{3p}$ is the difference of the two-point differentiations. 
The three-point demodulation can doubly suppress the low-frequency noise, i.e., 
it works as a second-order high pass filter as shown in Fig.~\ref{fig-tf}. 
Therefore, we can suppress the residual line noise in the two-point demodulation by using the 
three-point demodulation 
without study for the line noise sources.  
It can be explained intuitively that
smoothing with three neighboring points is more powerful for the low-frequency noise than the smoothing with 
two neighboring points.
The extracted polarization signals do not change unless the demodulation frequency is different from the modulation frequency.
%

\begin{figure}[htb]
\centering
\includegraphics[width=7.5cm]{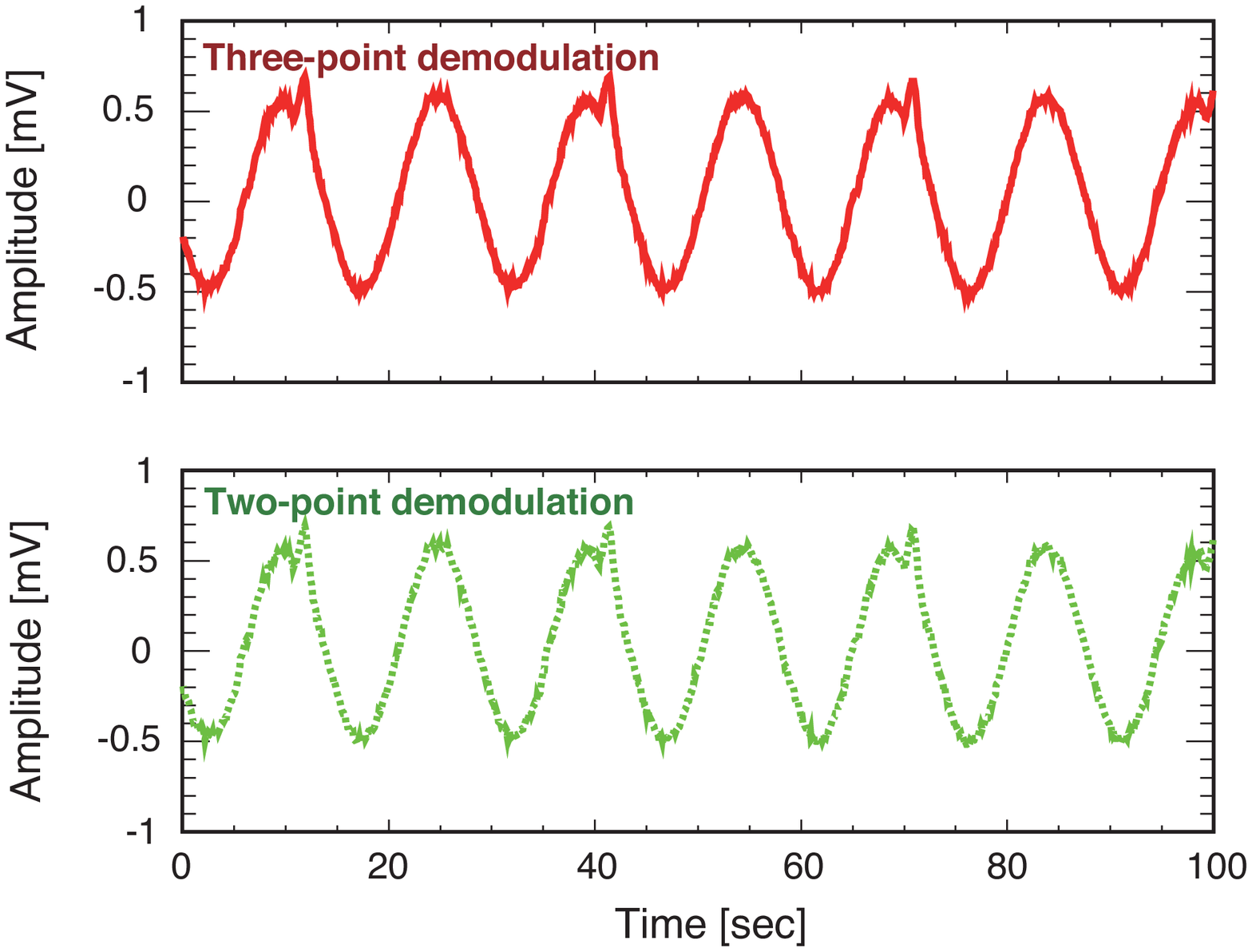}
\caption{
Measured polarization signals with three-point demodulation (top panel) and with two-point demodulation (bottom panel).
Both lock-in amplitudes are consistent with $0.03\%$ accuracy. 
}\label{data_fig0}
\centering
\includegraphics[width=7.5cm]{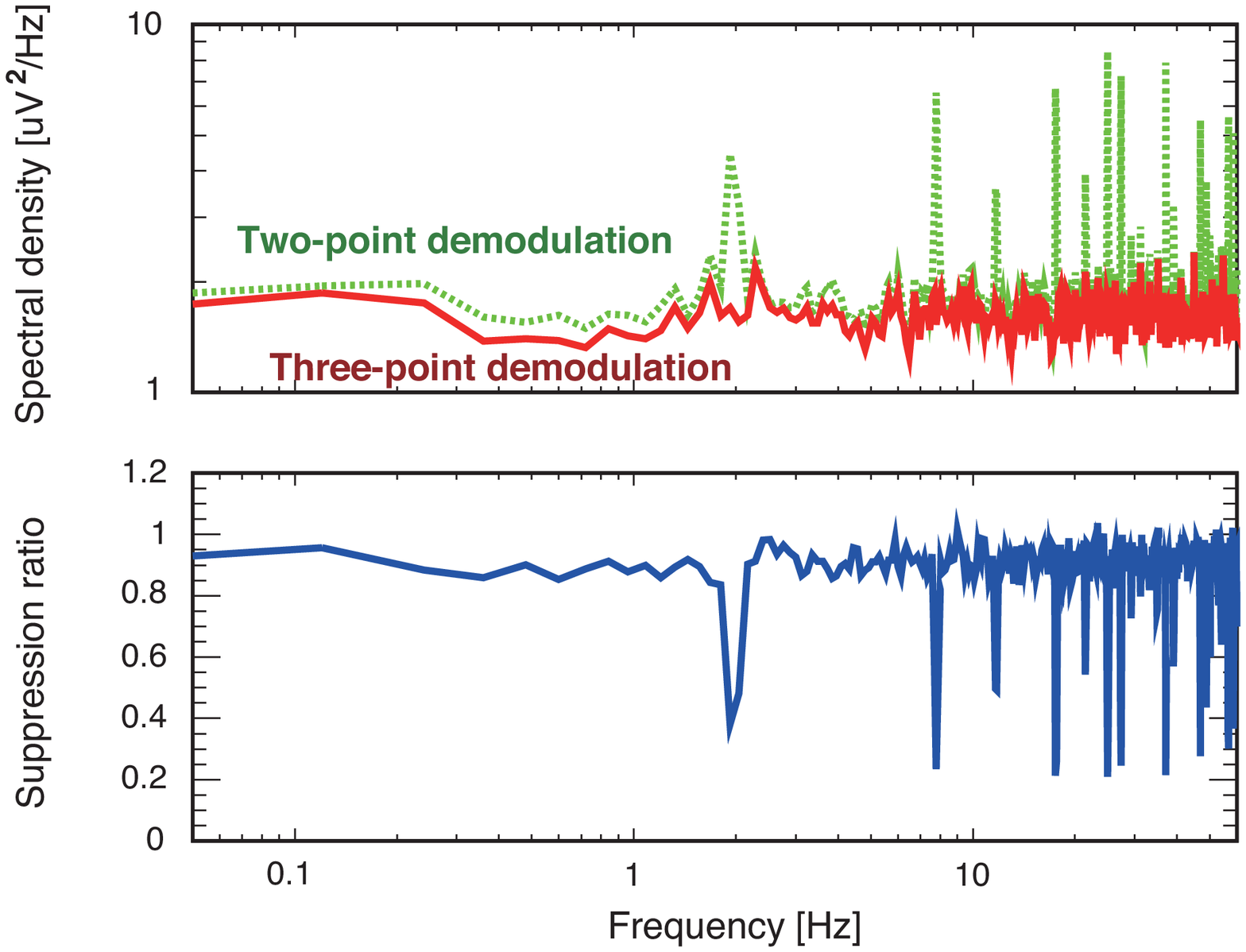}
 \caption{
Top panel shows noise spectral densities with two-point demodulation (dotted curves) and three-point demodulation (solid curves).
Their ratio as a function of frequency is shown in bottom panel.
Most of line noise is strongly suppressed by the three-point demodulation.
The improvement of the noise floor level ($8.4\%$) is also confirmed. 
}\label{data_fig2}
\end{figure}

We evaluate the actual impact of the three-point demodulation in real data.
The demodulation functions are implemented on the readout electronics board~\cite{adc} to be quickly used in the experiment.
The details of the experimental setup are described in~\cite{hase}.
Under the condition of 14 Kelvin of load temperature condition, 
we irradiate 0.6 Kelvin of the polarization which is rotated at $\sim30$ seconds of cycle;
the QUIET's detector~\cite{QUIET} 
measures sinusoidal polarization response in the demodulated data (Fig.~\ref{data_fig0}). 
We confirmed both demodulation methods measure the same polarization signals; both lock-in amplitudes 
are consistent within the precision of $0.03\%$.

We also measure 
the demodulated data without any variation of the input polarization signals.
The top panel of Fig.~\ref{data_fig2} shows noise spectral densities of each demodulation. 
Their ratio is also shown in the bottom panel of Fig.~\ref{data_fig2}. 
Most of line noise disappeared as expected. 
We also found 8.4\% of improvement for the noise floor in the demodulated data. 
This is the by-product of the three-point demodulation. 
The measured noise floor is determined by the input noise level at around the modulation frequency. 
By using the two-point demodulation, 
contamination from the $1/f$ noise has not been suppressed perfectly. 
The three-point demodulation improves the suppression.

To understand the improvement of the noise floor, we performed a simulation. 
We generated the noise stream that corresponded to the second term in Eq.~(\ref{eq1}) 
based on the simple model that consists of the white noise and the $1/f$ noise as $N (1 + (f_{\rm knee}/f)^{\alpha})$, where 
$f_{\rm knee}$ is the knee frequency, $\alpha$ is index to model the shape empirically, 
and $N$ is the white noise level in a spectral density. 
The generated noise stream also includes a line noise at 20~Hz. 

\begin{figure}[htb]
\centering
\includegraphics[width=7.5cm]{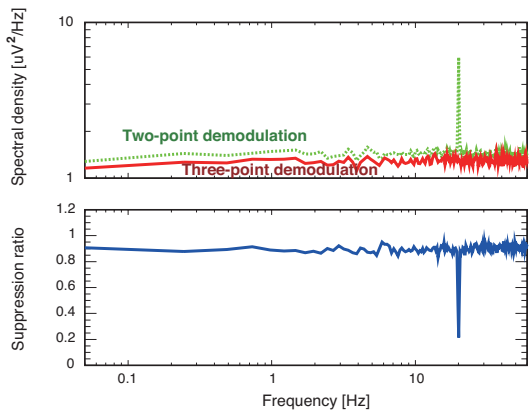}
\caption{
Noise spectral densities with two-point and three-point demodulation for simulation data (top panel).
Their ratio as a function of frequency (bottom panel). 
We use ($f_{\rm knee}$, $\alpha$, $N$)=(2~kHz, 1.5, 1.0~$\mu$V$^2$/Hz) for for the input parameters. 
Line noise at 20~Hz are strongly suppressed by the three-point demodulation.
The improvement of the noise floor level is also found. 
}\label{sim_fig1}
\end{figure}
\begin{table}[htb]
 \caption{
Simulated noise floor for each demodulation scheme: two-point demodulation ($2p$), three-point demodulation ($3p$), 
four-point demodulation ($4p$). 
Input noise properties are based on the ranges of QUIET's detector. 
On average, several percent improvement is expected by changing two-point to three-point demodulation. 
The improvement from three-point to four-point is at most 1\%. }\label{my_tab}
  \begin{tabular}{
@{\hspace{0.2cm}}
  c
@{\hspace{0.3cm}}
  c
@{\hspace{0.3cm}}  
  c
@{\hspace{0.5cm}}  
  c
@{\hspace{0.5cm}}  
  c
@{\hspace{0.3cm}}  
  c
@{\hspace{0.3cm}}
  }
   \hline 
   \multicolumn{3}{c}{Input noise properties} & \multicolumn{3}{c}{Noise floor~($\mu$V$^2$/Hz)} \\ 
    $f_{\rm knee}$~(kHz) & $\alpha$  & $N$~($\mu$V$^2$/Hz) & $2p$ & $3p$ & $4p$ \\ \hline
   1.0 &  1.0 &  1.0 &  1.18 & 1.16 & 1.15\\
   1.0 &  1.5 &  1.0 &  1.21 & 1.15 & 1.13\\
   2.0 &  1.0 &  1.0 &  1.31 & 1.27 & 1.26\\
   2.0 &  1.5 &  1.0 &  1.47 & 1.32 & 1.30\\ 
\hline
  \end{tabular}
\end{table}
Figure~\ref{sim_fig1} shows the noise spectral densities from both demodulations for the simulation data 
with ($f_{\rm knee}$, $\alpha$, $N$)=(2.0~kHz, 1.5, 1.0~$\mu$V$^2$/Hz) and their ratio as a function of frequency.  
We confirmed both the line suppression and the improvement of the noise floor by 10.1\%. 
The noise floor depends on the noise properties as listed in Table~I. 
In case of the higher $f_{\rm knee}$ or higher $\alpha$, the noise floor in the two-point demodulation is higher than 
one for the three-point. 
In other words, the three-point demodulation has higher suppression power for the $1/f$ noise contamination at around the modulation frequency. 
Typical detector properties of QUIET's detectors are ($f_{\rm knee}$, $\alpha$, $N$) is ($1$--$2$~kHz, $1$--$1.5$,  1.0~$\mu$V$^2$/Hz).
We expect several percent of improvement for the noise level in the demodulated data. 
This simulation explains the observed improvements of the noise floor in the real data.

We also estimate the level of the noise floor with four-point demodulation. 
The difference from the three-point demodulation is at most a few percents (Tab.~\ref{my_tab}). 
Compared with the complexity of implementing the logic in the readout electronics, 
its impact is relatively low.
We could achieve sufficient performance with the three-point demodulation within the range of the 
parameters for the detectors.

In summary, we suggest using three-point demodulation for coherent detectors in the CMB experiments. 
This innovative demodulation strongly suppresses non-white noise, e.g. the line noise, in the data,
which is one of the most important issues for CMB polarization experiments. 
The three-point demodulation is also immune to the tail of $1/f$ noise, 
which results in improvement of the measured noise floor. 
We confirmed both impacts in the real data by using QUIET's detector. 
This technique is applicable to ongoing experiments, e.g., PLANK-LFI, and proposed experiments, e.g., QUIET Phase-II.

We acknowledge QUIET collaboration, 
especially California Institute of Technology, Jet Propulsion Laboratory
and Fermi National Accelerator Laboratory for providing detectors and their electronics to control the detector bias.
We also thank Donna L. Kubik for carefully reading the manuscript. 
This work was supported by a Grant-in-Aid for Scientific Research on Innovative Areas (21111002) of The Ministry of Education, Culture, Sports, Science, and Technology, Japan.


\begin{thebibliography}{0}%
\makeatletter
\providecommand \@ifxundefined [1]{%
 \@ifx{#1\undefined}
}%
\providecommand \@ifnum [1]{%
 \ifnum #1\expandafter \@firstoftwo
 \else \expandafter \@secondoftwo
 \fi
}%
\providecommand \@ifx [1]{%
 \ifx #1\expandafter \@firstoftwo
 \else \expandafter \@secondoftwo
 \fi
}%
\providecommand \natexlab [1]{#1}%
\providecommand \enquote  [1]{``#1''}%
\providecommand \bibnamefont  [1]{#1}%
\providecommand \bibfnamefont [1]{#1}%
\providecommand \citenamefont [1]{#1}%
\providecommand \href@noop [0]{\@secondoftwo}%
\providecommand \href [0]{\begingroup \@sanitize@url \@href}%
\providecommand \@href[1]{\@@startlink{#1}\@@href}%
\providecommand \@@href[1]{\endgroup#1\@@endlink}%
\providecommand \@sanitize@url [0]{\catcode `\\12\catcode `\$12\catcode
  `\&12\catcode `\#12\catcode `\^12\catcode `\_12\catcode `\%12\relax}%
\providecommand \@@startlink[1]{}%
\providecommand \@@endlink[0]{}%
\providecommand \url  [0]{\begingroup\@sanitize@url \@url }%
\providecommand \@url [1]{\endgroup\@href {#1}{\urlprefix }}%
\providecommand \urlprefix  [0]{URL }%
\providecommand \Eprint [0]{\href }%
\providecommand \doibase [0]{http://dx.doi.org/}%
\providecommand \selectlanguage [0]{\@gobble}%
\providecommand \bibinfo  [0]{\@secondoftwo}%
\providecommand \bibfield  [0]{\@secondoftwo}%
\providecommand \translation [1]{[#1]}%
\providecommand \BibitemOpen [0]{}%
\providecommand \bibitemStop [0]{}%
\providecommand \bibitemNoStop [0]{.\EOS\space}%
\providecommand \EOS [0]{\spacefactor3000\relax}%
\providecommand \BibitemShut  [1]{\csname bibitem#1\endcsname}%
\let\auto@bib@innerbib\@empty
\end{thebibliography}%


\begin{thebibliography}{99}%

\bibitem{intro1} L. M. Krauss, S. Dodelson, and S. Meyrt, Science {\bf 328}, 989-992 (2010).

\bibitem{Durrer:2008eb} R.~Durrer, {\it THE COSMIC MICROWAVE BACKGROUND} (Cambridge Unisersity Press, 2008).

\bibitem{QUIET} QUIET Collaboration, Astrophys. J. {\bf 741}, 111 (2011).

\bibitem{PLANK} A. Mennella, M. Bersanelli, R. Butler, A. Curto, F. Cut-
taia, {\it et al}., Astronomy \& Astrophysics {\bf 536}, A4 (2011).


\bibitem{footnote} The modulation/demodulation at a higher frequency is a
good scheme to suppress the non-white noises. However, such modulation 
makes more the phase flips on the phase switches. The phase flips 
induce ringing spikes which must be masked. Compared with the sampling 
loss induced by the mask, the modulation frequency of 4~kHz is a natural 
condition to maximize the power of the noise suppression.

\bibitem{adc} K. Ishidoshiro, M. Nagai, T. Higuchi, M. Hasegawa, M. Hazumi, 
M. Ikeno, O. Tajima, M. Tanaka, and T. Uchida, 
To appear in IEEE Transactions on Nuclear Science (TNS) (2012), 
arXiv:1112.0644 [astro-ph.IM]. 

\bibitem{hase} M. Hasegawa, O. Tajima, Y. Chinone, M. Hazumi, K. Ishidoshiro, 
and M. Nagai, Rev. Sci. Instrum. {\bf 82}, 054501 (2011).

\end{thebibliography}
\end{document}